\let\accentvec\vec 
\RequirePackage{fix-cm}
\documentclass[smallcondensed]{svjour3}     
\let\vec\accentvec
\usepackage{graphicx}
\usepackage{dcolumn}
\usepackage{bm}
\usepackage{amssymb,amsfonts,amsmath}
\usepackage{makecell}
\usepackage{braket} 
\usepackage{enumerate}
\usepackage{soul}
\usepackage{xcolor}
\usepackage{color}
\usepackage{tcolorbox}
\usepackage{diagbox}

\soulregister\cite7 
\soulregister\citep7 
\soulregister\citet7 
\soulregister\ref7 
\soulregister\pageref7 
\journalname{Journal name to insert}
\begin{document}
\title{Hash function based on controlled alternate quantum walks with memory
}


\author{Qing Zhou         \and
        Songfeng Lu 
}
\institute{Qing Zhou \at
              School of Cyber Science and Engineering, Huazhong University of Science and Technology, Wuhan 430074, China \\
           \and
           Songfeng Lu \at
              School of Cyber Science and Engineering, Huazhong University of Science and Technology, Wuhan 430074, China \\
            Corresponding author,   \email{lusongfeng@hotmail.com}             \\
}
\date{Received: date / Accepted: date}

\maketitle

\begin{abstract}
We propose a new hash function QHFM based on controlled alternate quantum walks with memory on cycles, where the $j$th message bit decides whether to run quantum walk with one-step memory or to run quantum walk with two-step memory at the $j$th time step, and the hash value is calculated from the resulting probability distribution of the walker. Numerical simulation shows that the proposed hash function has near-ideal statistical performance and is at least on a par with the state-of-the-art hash functions based on quantum walks in terms of sensitivity of hash value to message, diffusion and confusion properties, uniform distribution property, and collision resistance property; and theoretical analysis indicates that the time and space complexity of the new scheme are not greater than those of its peers. The good performance of QHFM suggests that quantum walks that differ not only in coin operators but also in memory lengths can be combined to build good hash functions, which, in turn, enriches the construction of controlled alternate quantum walks.
\keywords{Hash function \and Quantum walks with memory \and Controlled alternate quantum walks \and Statistical properties \and time and space complexity}
\end{abstract}

\section{Introduction}\label{sec:1}
As one of the principal tools of information security, cryptographic hash functions not only act as essential components of identification, message authentication, digital signatures, and random number generation, but also play an important part in privacy amplification process of quantum key distribution~\cite{Bennett1995PrivAmplif}. Classical hash functions based on hard computational problems are, however, subject to an inherent security limitation: the existence of one-way functions is still an open conjecture that cannot be proved (a proof, with no assumptions, of existence would establish \textbf{P}$\ne$\textbf{NP}~\cite{Menezes1996Book}). As a result, they only satisfy computational security and are challenged by cryptanalysis equipped with quantum algorithms. Such a fact stimulates researchers to develop hash functions with a higher level of security, such as hash functions based on (or inspired by) quantum computing~\cite{Ablayev2016qHash,Ablayev2016Balanced,Vasiliev2016qHash,Ziatdinov2016qHash,Ablayev2013qHash,Ablayev2020uniHash,Yang2021qwHash,Yang2019qwHash,Yang2018qwHash264,Yang18qwHash221,Li2018qwHash,Cao2018qwHash,Yang2016qHash,Li2013qwHash}, whose preimage resistance property is ensured by quantum mechanics rather than hardness assumptions.

There are two kinds of quantum-computing-based hash functions: classical-quantum hash functions based on quantum one-way functions~\cite{Ablayev2016qHash,Ablayev2016Balanced,Vasiliev2016qHash,Ziatdinov2016qHash,Ablayev2013qHash,Ablayev2020uniHash} (hereafter, simply QOWF-based hash functions) and classical-classical hash functions based on discrete quantum walks on cycles~\cite{Yang2021qwHash,Yang2019qwHash,Yang2018qwHash264,Yang18qwHash221,Li2018qwHash,Cao2018qwHash,Yang2016qHash,Li2013qwHash} (hereafter, simply QW-based hash functions). The former have balanced one-way resistance property and collision resistance property that are well-defined and strictly proved in the quantum setting, the latter take advantage of the chaotic characteristics of quantum walks and belong to dedicated hash functions, whose capabilities of collision resistance are difficult to prove and are mainly assessed by means of statistical analysis. On the other hand, of QOWF-based hash functions, the output length and the number of hashing parameters are both positively correlated with the input size, while QW-based hash functions map messages of arbitrary finite length to digests of fixed length. In addition, the output length of QW-based hash functions can be easily extended (to withstand brute-force attacks) by increasing the number of nodes of the cycle or the number of hash bits ``contributed'' by each node, and the hash result can be calculated classically. Thus, QW-based hash functions are of greater practical utility before large-scale quantum computers are built, for they can currently be used to improve the security of hash-function-based schemes.

The essence of the design of QW-based hash functions is combining two or more different quantum walk procedures governed by evolution operators $\{U_0,U_1,\dots\}$ to construct a controlled alternate quantum walk (CAQW) model, where the choice among $\{U_0,U_1,\dots\}$ at the $j$th time step is determined by the $j$th bit of a binary string. Theoretically, a valid CAQW model could be constructed if the walker can ''switch'' freely among $\{U_0,U_1,\dots\}$, and evolution operators that only differ in coin transform naturally satisfy this requirement. Therefore, various quantum walks, such as one-dimensional broken-line quantum walks~\cite{Yang2021qwHash}, one-dimensional one-particle quantum walks~\cite {Yang2019qwHash,Yang2018qwHash264,Yang18qwHash221}, two-dimensional one-particle quantum walks~\cite{Li2018qwHash}, quantum walks on Johnson graphs~\cite{Cao2018qwHash}, one-dimensional two-particle (interacting) quantum walks~\cite{Yang2016qHash,Li2013qwHash}, and one-dimensional quantum walks with memory (QWM)~\cite{Li2020QWM,Dai2020QWM,Zhou2019QW2M,Li2016QWM,Gettrick2014QWM,Gettrick2010QW1M} can all be used to construct valid (but may not good) hash functions as long as they are modified to utilize coin operators controlled by input messages. Among these walks, the evolution of QWM is governed by three (rather than two) stages: flipping a coin, determining the next direction according to the coin state and the previous direction(s), and moving a step according to the new direction. Here, an extra operator---the direction-determine transform---could be taken into account when designing hash functions based on QWM: the alternately performed evolution operators can differ in coin transform or direction-determine transform, or both.

To examine the feasibility and utility of this idea, we combine two quantum walks with different memory lengths, i.e., QW1M~\cite{Gettrick2010QW1M} and QW2M~\cite{Zhou2019QW2M}, to achieve a valid CAQW process, for different quantum walks with unequal memory lengths typically have different direction-determine transforms, and they can also use different coin operators. Based on this walking process, we construct a new hash function (named QHFM) and then assess its performance. Simulation results show that the statistical properties of the proposed hash function are as good as those of the existing QW-based hash functions, and theoretical analysis indicates that the space and time complexity of QHFM are not greater than those of theirs peers.

\section{One-dimensional Controlled Quantum Walk with One- and Two-step Memory}\label{sec:2}
A one-dimensional controlled quantum walk with one- and two-step memory (CQWM) takes place in the Hilbert space $\mathcal{H}_p\otimes\mathcal{H}_{dr_2}\otimes\mathcal{H}_{dr_1}\otimes\mathcal{H}_c$ spanned by vectors $\Ket{x,dr_2,dr_1,c}$, where $c$ (with $c\in\{0,1\}$, or $c\in\mathbb{Z}_2$) is the coin state, $dr_1$ (with $dr_1\in\mathbb{Z}_2$) is the direction of the most recent step (0 stands for left and 1 stands for right), $dr_2$ (with $dr_2\in\mathbb{Z}_2$) is the direction of the penultimate step, and $x$ is the current position. If the walker moves on a line, then $x\in\mathbb{Z}$ (all integers); and if the walker moves on a cycle with $n$ nodes, then $x\in\{0,1,2,\ldots,n-1\}$ (or $x\in\mathbb{Z}_n$).

Formally, the evolution of CQWM controlled by a $t$-bit string $msg=$\\$(m_1,m_2,\dots,m_t)\in\{0,1\}^t$ is the product of $t$ unitary transforms
\begin{equation}\label{eq:1}
U_{msg}=U^{(m_t)}U^{(m_{t-1})}\cdots U^{(m_2)}U^{(m_1)},
\end{equation}
Here $U^{(m_j)}$ ($1\leq j\leq t$) is the one-step transform controlled by the $j$th bit of $msg$, and it is defined as
\begin{equation}\label{eq:2}
U^{(m_j)}=S\cdot \left(I_n\otimes D^{(m_j)}\right)\cdot \left(I_{4n}\otimes C^{(m_j)}\right),
\end{equation}
where $C^{(m_j)}$ is a $2\times 2$ coin operator controlled by $m_j$, $I_k$ ($k=4n$ or $n$) is a $k\times k$ identity operator, $D^{(m_j)}$ is an $8\times 8$ direction-determine operator controlled by $m_j$, and $S$ is the conditional shift operator controlled by the next direction. If $m_j=0$, then $C^{(m_j)}$ is parameterized by an angle $\theta_0$, i.e.,
\begin{equation}
C^{(0)}=
\begin{pmatrix}\label{eq:3}
\text{cos}(\theta_0) & \text{sin}(\theta_0) \\
\text{sin}(\theta_0) & -\text{cos}(\theta_0)
\end{pmatrix},
\end{equation}
and $D^{(m_j)}$ (becomes $D^{(0)}$) describes the direction-determine process of QW1M; if $m_j=1$, then $C^{(m_j)}$ is parameterized by another angle $\theta_1$, i.e.,
\begin{equation}\label{eq:4}
C^{(1)}=
\begin{pmatrix}
\text{cos}(\theta_1) & \text{sin}(\theta_1) \\
\text{sin}(\theta_1) & -\text{cos}(\theta_1)
\end{pmatrix},
\end{equation}
and $D^{(1)}$ describes the direction-determine process of QW2M.

The direction-determine transforms of QW1M~\cite{Gettrick2010QW1M} and QW2M~\cite{Zhou2019QW2M} can be respectively written as $DT_0:\Ket{dr_1,c}\to \Ket{dr_1\oplus\bar{c},c}$ and $DT_1:\Ket{dr_2,dr_1,c}\to \Ket{dr_1,dr_2\oplus \bar{c},c}$, where $\bar{c}\equiv 1\oplus c$, $dr_1 \oplus \bar{c}$ in the first expression specifies the next direction of the walker performing QW1M, and $dr_2\oplus \bar{c}$ in the second expression specifies the next direction of the walker performing QW2M. To enable QW1M and QW2M to be performed alternately, one may add a redundant state $\Ket{dr_2}$ into QW1M and let $D^{(0)}:\Ket{dr_2,dr_1,c}\to \Ket{dr_2,dr_1\oplus \bar{c},c}$ determines the next direction when the controlling bit equals 0; otherwise, the next direction is determined by $D^{(1)}=DT_1$.

Following Ref.~\cite{Zhou2019QW2M}, any 4-term basis state $\Ket{x, dr_2,dr_1,c}$ in $\mathcal{H}_p\otimes\mathcal{H}_{dr_2}\otimes\mathcal{H}_{dr_1}\otimes\mathcal{H}_c$ can be rewritten as a 2-term basis state $\Ket{x,j}=\Ket{x, 2^2dr_1+2^1 dr_2+2^0 c}$ in $\mathcal{H}_p\otimes\mathcal{H}^8$, where $\mathcal{H}^8$ is the 8-dimensional Hilbert space. conversely, from any 2-term basis state $\Ket{x,j}$ ($j\in\mathbb{Z}_8$), one can deduce the coin value and the most recent two directions as follows:
\begin{equation}\label{eq:5}
\begin{cases}
    c=j\bmod 2, \\
    dr_2=(j\bmod 4-j\bmod 2)/2, \\
    dr_1=(j-j\bmod 4)/4.
\end{cases}
\end{equation}
According to this correspondence, $D^{(0)}$ can be reformulated to
\begin{displaymath}
\Ket{2^2dr_1+2^1dr_2+2^0c}\to\Ket{2^2(dr_1\oplus\bar{c})+2^1dr_2+2^0c},
\end{displaymath}
or, under the 2-term states
\begin{equation}\label{eq:6}
    D^{(0)}:\Ket{j}\to\Ket{4\left[(j-j\bmod 4)/4\oplus (j\bmod 2)\oplus 1\right] + j\bmod 4}.
\end{equation}
Analogously, $D^{(1)}$ can be expressed as
\begin{equation}\label{eq:7}
D^{(1)}:\Ket{j}\to\Ket{4\left[(j\bmod 4-j\bmod 2)/2\oplus (j\bmod 2)\oplus 1\right] + (j-j\bmod 4)/2+j\bmod 2}.
\end{equation}
With formulas~(\ref{eq:6}) and~(\ref{eq:7}), one can verify that $D^{(0)}$ and $D^{(1)}$ are both unitary.

Once the next direction, the new $dr_1$, is determined, the walker then moves according to the shift operator controlled by $dr_1$. If the walk takes place on a line, then the action of $S$ is expressed as $\Ket{x,dr_2,dr_1,c} \to \Ket{x+2dr_1-1,dr_2,dr_1,c}$; if the walk takes place on a cycle with $n$ nodes, then $S$ becomes $\Ket{x,dr_2,dr_1,c} \to \Ket{x+2dr_1-1\pmod{n},dr_2,dr_1,c}$, which can be reformulated (in 2-term states) to
\begin{equation}\label{eq:8}
S:\Ket{x,j} \to \Ket{x+(j-j\bmod 4)/2-1\pmod{n},j}.
\end{equation}
In formula~(\ref{eq:8}), the next position is calculated using modular arithmetic under modulus $n$.

\section{Hash Function Using Quantum Walks with One- and Two-step Memory on Cycles}
The proposed hash function is constructed by running CQWM on a circle with $n$ nodes under the control of the input message $msg$, where each node contributes $m$ bits to the hash result $H(msg)$. The process of CQWM-based hash function is described as follows:
\begin{enumerate}
\item[(1)] Select the values of parameters $(n,m,l,\theta_0,\theta_1,\alpha)$ satisfying the following constraints: $n$ is odd; $n\times m$ equals the bit length of the hash value; $10^l\gg 2^m$; and $\theta_0,\theta_1,\alpha\in(0,\pi/2)$.
\item[(2)] Initialize the walker in the state $\Ket{\psi_0}=\text{cos}\alpha\Ket{0,1,0,0}+\text{sin}\alpha\Ket{0,1,0,1}$ (or, in the 2-term state $\Ket{\psi_0}=\text{cos}\alpha\Ket{0,2}+\text{sin}\alpha\Ket{0,3}$).
\item[(3)] Apply $U_{msg}$ to $\Ket{\psi_0}$ and generate the resulting probability distribution $prob=$\\$(p_0,p_1,\dots,p_{n-1})$, where $p_x$ ($x\in \mathbb{Z}_n$) is the probability that the particle locates at node $x$ when the walk is finished.
\item[(4)] The hash value of $msg$ is a sequence of $n$ blocks $H(msg)=B_0\lVert B_1\lVert \dots \lVert B_{n-1}$, where each block $B_x$ is the $m$-bit binary representation of $\lfloor p_x\cdot 10^l\rfloor\bmod 2^m$ ($\lfloor\cdot\rfloor$ denotes the floor of a number), and $B_x\lVert B_{x+1}$ denotes the concatenation of $B_x$ and $B_{x+1}$.
\end{enumerate}

\section{Statistical Performance Analysis}\label{sec:4}
QHFM, like other QW-based hash functions~\cite{Yang2021qwHash,Yang2019qwHash,Yang2018qwHash264,Yang18qwHash221,Li2018qwHash,Cao2018qwHash,Yang2016qHash,Li2013qwHash}, belongs to dedicated hash functions, whose performances are mainly evaluated through statistical analysis. To make our statistical tests reusable and usable by anyone else, we perform these tests on a collection of items (i.e., input messages) randomly drawn from an open dataset, named ``arXiv Dataset'', of about 1.8 million records and upload the complete MATLAB code for hash tests to ``GitHub''. See the dataset at https://www.kaggle.com/Cornell-University/arxiv and the test code at \\ https://github.com/Chloe-Zhouqing/Hash-functions-based-on-quantum-walks.

To make comparisons between the proposed scheme and the existing ones with (detailed) experimental results~\cite{Yang2021qwHash,Yang2019qwHash,Yang2018qwHash264,Yang18qwHash221,Li2018qwHash,Cao2018qwHash,Yang2016qHash} in a fair and informative manner, we consider seven ``instances'' QHFM-$L$ ($L\in\{296,264,221,200,195,136,120\}$) of QHFM, where QHFM-$L$ produces $L$-bit hash values and will be compared with the existing QW-based hash functions with $L$- or close-to-$L$-bit output length (QHFM-136 and QHFM-120 will be compared with the 128-bit scheme in Ref.~\cite{Yang2016qHash}). Different instances of QHFM share the same $l$ values, same $\theta_0$ values, same $\theta_1$ values, and the same $\alpha$ values, which are taken to be 8, $\pi/4$, $\pi/3$, and $\pi/4$, respectively. Distinction between QHFM-$L$ and QHFM-$L'$ ($L\neq L'$) lies in the values of $n$ and $m$, which are listed in Table~\ref{tab:table1}.

\begin{table}
\caption{Values of parameters chosen for the seven instances of the proposed hash scheme}\label{tab:table1}
\begin{tabular}{crr}
\hline\noalign{\smallskip}
Hash Instances & $n$ & $m$ \\
\hline\noalign{\smallskip}
    QHFM-296 & 37 & 8  \\
    QHFM-264 & 33 & 8  \\
    QHFM-221 & 17 & 13 \\
    QHFM-200 & 25 & 8  \\
    QHFM-195 & 15 & 13 \\
    QHFM-136 & 17 & 8  \\
    QHFM-120 & 15 & 8  \\
\hline
\end{tabular}
\end{table}
\subsection{Sensitivity of Hash Value to Message}\label{sec:4.1}
Let $msg_0$ be an original message and $msg_j$ ($j\in\{1,2,3\}$) the slightly modified result of $msg_0$, which are obtained under the following four conditions:
\begin{itemize}
    \item[] Condition 1: Randomly choose an original message $msg_0$;
    \item[] Condition 2: Flip a bit of $msg_0$ at a random position and then obtain the modified message $msg_1$;
    \item[] Condition 3: Insert a random bit into $msg_0$ at a random position and then obtain $msg_2$;
    \item[] Condition 4: Delete a bit from $msg_0$ at a random position and then obtain $msg_3$.
\end{itemize}

The sensitivity of hash value to message is assessed by comparing the hash values $H(msg_j)$ of the modified messages with the hash value $H(msg_0)$ of the original one. In our sensitivity test, a record is randomly picked out from the arXiv Dataset, then the article abstract within this record serves as $msg_0$.

Corresponding to the conditions above, four hash values in hexadecimal format produced by QHFM-195 are obtained as follows:
\begin{itemize}
    \item[] Condition 1: $H(msg_0)=$``3 5A 2B 76 96 74 1C F7 51 09 2E AB 1F CB 6A C0 33 77 46 61 E5 D1 E4 38 EC'';
    \item[] Condition 2: $H(msg_1)=$``4 BC EC C7 0E A9 2B 5C 5C 93 34 30 69 E9 3A EC 1B D3 D3 95 7B 0F DF 5A 31'';
    \item[] Condition 3: $H(msg_2)=$``5 00 5C 40 AB AB 2F 26 9B AB D7 AF B5 23 4F 16 20 5C 63 A0 30 6D 5E 0C 15'';
    \item[] Condition 4: $H(msg_3)=$``0 5D 14 81 F1 29 CB E7 BE CB 01 F6 53 48 E8 90 D4 CD 35 C3 C7 55 DB 80 E8''.
\end{itemize}

Notice that the the first hexadecimal digit of the hash value under condition $j$ only represents the first three (rather than four) bits of $H(msg_j)$, since the output length of QHFM-195 is not a multiple of four.

\begin{figure}
    \centering
    \includegraphics[width=0.75\columnwidth]{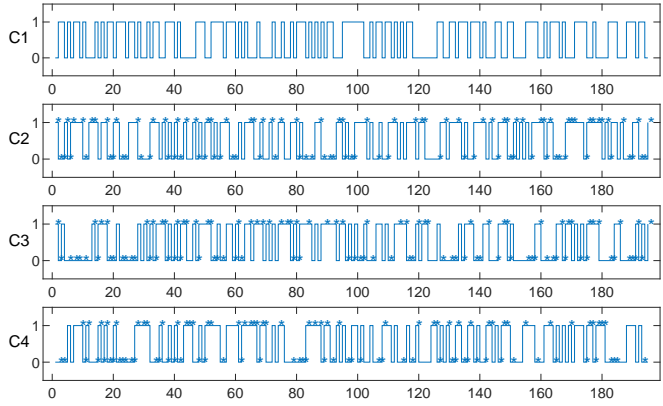}
    \caption{(Color online) Plots of the hash values produced by QHFM-195 under the four conditions, where $\text{C}j$ stands for Condition $j$ ($j\in\{1,2,3,4\}$). In the $j$th subgraph with $j>1$, each asterisk (*) marks a different bit between $H(msg_{j-1})$ and $H(msg_0)$.}
    \label{fig:fig1}
\end{figure}

The plots of hash values $H(msg_0)$, $H(msg_1)$, $H(msg_2)$, and $H(msg_1)$ in binary format are shown in Fig.~\ref{fig:fig1}, which indicates that a tiny modification to the message could cause a significant change in the hash value, and the positions of those changed bits are evenly distributed over the entire interval $[1,195]$ of position numbers. A similar result can be obtained using any other instance of QHFM; thus, the output digest of the proposed hash scheme is highly sensitive to its input message.
\subsection{Diffusion and Confusion Properties}\label{sec:4.2}
The test data for the diffusion and confusion properties of QHFM-$L$ is collected by making $N$ random draws (with replacement) from the arXiv Dataset. On each draw, an original message $msg_0$ is selected, then a slightly modified result $msg_1$ of this message is obtained by inverting a bit of $msg_0$ at a random position. Let $B_i$ be the Hamming distance between the hash values of the original and modified messages obtained on the $i$th draw and $N$ the number of draws, the diffusion and confusion properties (reflecting the avalanche effect) of the proposed hash instances are assessed based on the following four indicators:
\begin{itemize}
    \item[$\bullet$] mean changed bit number $\overline{B}=\begin{matrix} \sum_{i=1}^N B_i/N \end{matrix}$;
    \item[$\bullet$] mean changed probability $P=\left[\overline{B}/(n\times m)\right]\times 100\%$;
    \item[$\bullet$] standard deviation of the changed bit number\\
        $\Delta B=\sqrt{\left[1/(N-1)\right]\begin{matrix}\sum_{i=1}^N\left(B_i-\overline{B}\,\right)^2\end{matrix}}$;
    \item[$\bullet$] standard deviation of the changed probability\\
        $\Delta P=\sqrt{\left[1/(N-1)\right]\begin{matrix}\sum_{i=1}^N\left[B_i/(n\times m)-P\right]^2\end{matrix}}\times 100\%$.
\end{itemize}

The ideal values of $\overline{B}$ and $P$ are $(n\times m)/2$ and $50\%$, respectively; and smaller standard deviations ($\Delta B$ and $\Delta P$) are more desirable. For a specific hash function with fixed output length, $\overline{B}$ and $\Delta B$ are directly proportional to $P$ and $\Delta P$, respectively; thus, only $P$ and $\Delta P$, or a combination of them, e.g., $I_\text{DC}=(\Delta P+|P-50\%|)/2\times 100\%$, would suffice to assess the confusion and diffusion properties of this hash function: the smaller $I_\text{DC}$, the better the avalanche effect achieved. The diffusion and confusion test on QHFM-$L$ is performed with $N=10000$, and the simulation results are presented in Table~\ref{tab:table2}. For comparison, the reported results (with $N\ge 10000)$ of the corresponding variables for the existing QW-based hash schemes~\cite{Yang2021qwHash,Yang2019qwHash,Yang2018qwHash264,Yang18qwHash221,Li2018qwHash,Cao2018qwHash,Yang2016qHash} are also listed in the same table, where the values for Yang21-296 and Yang18-221 (which have multiple instances) are the test results for the representative instances of them. Here, the \textit{representative instance} of a hash scheme $X$ is the one that has the $P$ result closest to $50\%$ (over all instances of $X$).
\begin{table}
\caption{Diffusion-and-Confusion-Test Results for The Proposed and Existing QW-Based Hash Functions}\label{tab:table2}
\begin{tabular}{lrrrrr}
\hline\noalign{\smallskip}
Hash instances (Schemes) & $\overline{B}$~~~~~~& $P(\%)$~~& $\Delta B$~~& $\Delta P(\%)$ & $I_\text{DC}(\%)$ \\
\hline\noalign{\smallskip}
    QHFM-296 & 147.9101 & 49.9696 & 8.5997 & 2.9053 & 1.4679 \\
    QHFM-264 & 131.8667 & 49.9495 & 8.1378 & 3.0825 & 1.5665 \\
    QHFM-221 & 110.5313 & 50.0142 & 7.4455 & 3.3690 & 1.6916 \\
    QHFM-200 & 100.0205 & 50.0103 & 7.1654 & 3.5827 & 1.7965 \\
    QHFM-195 & 97.5591 & 50.0303 & 6.9844 & 3.5817 & 1.8060 \\
    QHFM-136 & 68.0530 & 50.0390 & 5.8782 & 4.3222 & 2.1806 \\
    QHFM-120 & 60.0914 & 50.0762 & 5.4699 & 4.5582 & 2.3172 \\
    Yang21-296~\cite{Yang2021qwHash} & 147.8640 & 49.9541 & 8.6141 & 2.9102 & 1.4781 \\
    Yang19-264~\cite{Yang2019qwHash} & 131.6803 & 49.8789 & 8.8877 & 3.3666 & 1.7439 \\
    Yang18-264~\cite{Yang2018qwHash264} & 132.1108 & 50.0420 & 8.0405 & 3.0457 & 1.5439 \\
    Yang18-221~\cite{Yang18qwHash221} & 112.7791 & 51.0313 & 8.2029 & 3.7117 & 2.3715 \\
    Li18-200~\cite{Li2018qwHash} & 99.9010 & 49.9505 & 7.1133 & 3.5567 & 1.8031 \\
    Cao18-195~\cite{Cao2018qwHash} & 124.7000 & 63.9600 & 6.4300 & 6.3000 & 10.1300 \\
    Yang16-128~\cite{Yang2016qHash} & 64.2894 & 50.2261 & 5.6686 & 4.4286 & 2.3274 \\
\hline
\end{tabular}
\end{table}

The values of $I_\text{DC}$ suggest that the test results for QHFM-264 is better than that for Yang19-264 but slightly poorer than that for Yang18-264, and the results for other instances of the proposed hash scheme are better than those for their peers (QHFM-296 vs. Yang21-296; QHFM-221 vs. Yang18-221; QHFM-200 vs. Li18-200; QHFM-195 vs. Cao18-195; QHFM-136 and QHFM-120 vs. Yang16-128). Thus, the diffusion and confusion properties of the proposed hash function outperform or are at least on a par with the existing QW-based hash schemes.
\subsection{Uniform Distribution Analysis}\label{sec:4.3}
Similar to the case of diffusion and confusion properties, the uniform distribution property (reflecting the strict avalanche effect) could also be assessed based on four indicators:
\begin{itemize}
    \item[$\bullet$] mean number of draws with flipped hash bit (over $n\times m$ bit positions)\\
        $\overline{T}=\begin{matrix} \sum_{j=1}^{n\times m} T_j/(n\times m) \end{matrix}$;
    \item[$\bullet$] mean percentage of draws with flipped hash bit $Q=\overline{T}/N\times 100\%$;
    \item[$\bullet$] standard deviation of the number of draws with flipped hash bit \\
        $\Delta T=\sqrt{\left[1/(n\times m-1)\right]\begin{matrix}\sum_{j=1}^{n\times m}(T_j-\overline{T})^2\end{matrix}}$;
    \item[$\bullet$] standard deviation of the percentage of draws with flipped hash bit\\
        $\Delta Q=\sqrt{\left[1/(n\times m-1)\right]\begin{matrix}\sum_{j=1}^{n\times m}\left[T_j/N-Q\right]^2\end{matrix}}\times 100\%$;
\end{itemize}
where $T_j$ ($j\in\{1,2,3,\dots ,n\times m\}$) is the number of draws on which a bit-flip occurs in the hash value at the $j$th bit position after a random message bit is inverted. The theoretical values of $\overline{T}$ and $Q$ are $(n\times m)/2$ and $50\%$, respectively.

Since $\overline{T}$ and $\Delta T$ are directly proportional to $Q$ and $\Delta Q$, respectively, the uniform distribution property of a hash function could be evaluated using $|Q-50\%|$ and $\Delta Q$: the smaller they are, the better the strict avalanche effect achieved. Additionally, the experimental value of $Q$ is always equivalent to the value of $P$ if the test data (i.e., $N$ pairs of original and modified messages) used in the diffusion and confusion test is re-used in the uniform distribution test. Such a result can also be obtained through a simple reasoning: in $P=\left(\begin{matrix}\sum_{i=1}^N B_i \end{matrix}\right)/(n\times m\times N)\times 100\%$ and $Q=\left(\begin{matrix}\sum_{j=1}^{n\times m} T_j \end{matrix}\right)/(n\times m\times N)\times 100\%$, both $\begin{matrix}\sum_{i=1}^N B_i \end{matrix}$ and $\begin{matrix}\sum_{j=1}^{n\times m} T_j \end{matrix}$ count the total number of hash bits that are flipped over $N$ draws. Thus, $T$ or $Q$ alone is insufficient for assessing the uniform distribution property of a hash function, it should be considered along with $\Delta Q$.

The uniform distribution test on QHFM-$L$ is conducted as follows:
\begin{enumerate}
    \item[(1)] Set $T_j=0$ for every bit position $j$ in the hash value.
    \item[(2)] Randomly draw an article record from arXiv Dataset, take the abstract of this article as the original message $msg_0$.
    \item[(3)] Randomly flip a bit of $msg_0$ and then generate the modified message $msg_1$.
    \item[(4)] Compute the hash values of the two messages and get the digest pair ($H(msg_0)$, $H(msg_1)$); compare $H(msg_0)$ with $H(msg_1)$ bit by bit, if $H(msg_0)$ differs from $H(msg_1)$ at the $j$th bit position, then the value of $T_j$ is incremented by one.
    \item[(5)] Repeat steps (2) to (4) $N$ times.
    \item[(6)] Calculate $\overline{T}$, $Q$, $\Delta T$, and $\Delta Q$ from the obtained data.
\end{enumerate}

The data collected in step (2) is re-used for different instances of the proposed hash scheme as well as for different hash properties (i.e., the diffusion and confusion properties, the uniform distribution property, and the collision resistant property). As a result, the experimental values of $P$ and $Q$ for each instances are equal, which gives $N\times P = \overline{T}$ and $|P-50\%|=|Q-50\%|$. On the other hand, for the existing schemes~\cite{Yang2021qwHash,Yang2019qwHash,Yang2018qwHash264,Yang18qwHash221,Li2018qwHash,Cao2018qwHash,Yang2016qHash}, the reported results of $\overline{T}$ are not equivalent to the corresponding outcomes of $N\times P$, this is probably because their input messages used in the uniform distribution test are not the same as that used in the diffusion and confusion test. Nevertheless, the reported values of $\overline{T}$ are generally close to the corresponding results of $N\times P$.

Since the test results of $\Delta Q$ (or $\Delta T$) for the existing schemes are unavailable for comparison, we collect reported data related to the uniform distribution property from Refs.~\cite{Yang2021qwHash,Yang2019qwHash,Yang2018qwHash264,Yang18qwHash221,Li2018qwHash,Cao2018qwHash,Yang2016qHash} as much as possible and list the corresponding results for the proposed and existing schemes in Table~\ref{tab:table3}, where ``****.**, same'' denotes a pair of identical values, and ``N/A'' means ``not available''. The values of $|Q-50\%|$ (in $\%$) for the existing schemes are deduced from the reported results of $\overline{T}$: $|Q-50\%|=|\overline{T}/N\times 100\%-50\%|$, where $N=16383$ for Cao18-195 and $N=10000$ for others. Similar to Table~\ref{tab:table2}, the values presented in the 8th and 11th rows are results for the representative instances of Yang21-296 and Yang18-221, respectively.
\begin{table}
\caption{Uniform-Distribution-Test Results for the Proposed and Existing QW-Based Hash Functions}\label{tab:table3}
\begin{tabular}{lrccr}
\hline\noalign{\smallskip}
\textrm{Instances (schemes)} & $N\times P,\overline{T}\quad\;$ &  $\Delta T$ & $\Delta Q(\%)$& $|P-50\%|,|Q-50\%|(\%)$\\
\hline\noalign{\smallskip}
    \ QHFM-296 & 4996.96, \,same\;\, &  48.4334 & 0.4843 & 0.0304, \,same$\ \qquad$\\
    \ QHFM-264 & 4994.95, \,same\;\, &  48.9253 & 0.4893 & 0.0505, \,same$\ \qquad$\\
    \ QHFM-221 & 5001.42, \,same\;\, &  51.6083 & 0.5161 & 0.0142, \,same$\ \qquad$\\
    \ QHFM-200 & 5001.03, \,same\;\, &  51.6897 & 0.5169 & 0.0103, \,same$\ \qquad$\\
    \ QHFM-195 & 5003.03, \,same\;\, &  50.5134 & 0.5051 & 0.0303, \,same$\ \qquad$\\
    \ QHFM-136 & 5003.90, \,same\;\, &  46.6002 & 0.4660 & 0.0390, \,same$\ \qquad$\\
    \ QHFM-120 & 5007.62, \,same\;\, &  48.6068 & 0.4861 & 0.0762, \,same$\ \qquad$\\
    \ Yang21-296~\cite{Yang2021qwHash} & 4995.41, 4998.1 &  N/A & N/A & 0.0459, 0.019$\ \qquad$\\
    \ Yang19-264~\cite{Yang2019qwHash} & 4987.89, 4996.6 &  N/A & N/A & 0.1211, 0.034$\ \qquad$\\
    \ Yang18-264~\cite{Yang2018qwHash264} & 5004.20, 5003.9 & N/A & N/A & 0.0420, 0.039$\ \qquad$\\
    \ Yang18-221~\cite{Yang18qwHash221} & 5103.13, N/A\ \,\; &  N/A & N/A & 1.0313, N/A$\,\ \qquad$\\
    \ Li18-200~\cite{Li2018qwHash} & 4995.05, 4998.2 &  N/A & N/A & 0.0495, 0.018$\ \qquad$\\
    \ Cao18-195~\cite{Cao2018qwHash} & 10478.57, 6495.0 & N/A & N/A & 13.9600,\,10.355$\ \qquad$\\
    \ Yang16-128~\cite{Yang2016qHash} & 5022.61, 4973.5 &  N/A & N/A & 0.2261, 0.265$\ \qquad$\\
\hline\noalign{\smallskip}
\end{tabular}
\end{table}

It can be seen from the last column of Table~\ref{tab:table3} that the experimental values of $P$ and $Q$ for QHFM-221, QHFM-195, and QHFM-136 together with QHFM-120 are closer to their theoretical values than those for Yang18-221, Cao18-195, and Yang16-128, respectively; and the values of $|Q-50\%|$ for the remaining instances of QHFM are on a par with those for their peers. In addition, the results of $\Delta Q$ for all instances of QHFM are very small, indicating that the proposed hash scheme has a very good uniform distribution property.

\begin{figure}
    \centering
    \includegraphics[width=0.75\columnwidth]{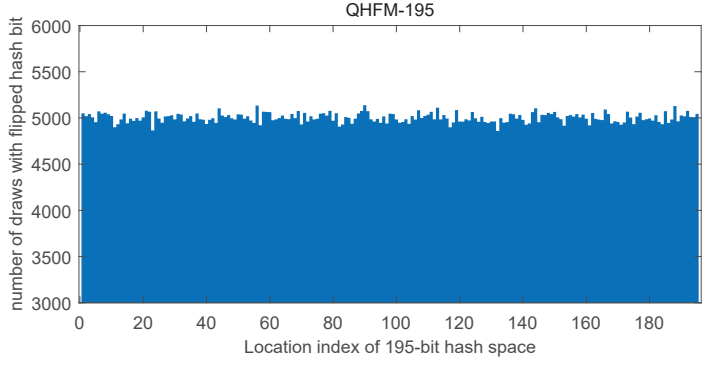}
    \caption{(Color online) Histogram of the 195-bit hash space, where $N=10000$.}\label{fig:fig2}
\end{figure}

To provide an intuitive description for this property of our scheme, we plot the number of draws with flipped hash bit on every bit position of QHFM-195 in Fig.~\ref{fig:fig2}, which suggests that the proposed scheme has a good resistance to statistical attacks.
\subsection{Collision Resistance}\label{sec:4.4}
The test data for the diffusion and confusion properties or the uniform distribution property can also be used to analyzing the collision resistance property, which is generally assessed in terms of two groups of indicators: (1) the number of draws $W_N^\text{\,e}(\omega)$ (out of $N$ random selections) on which the hash values of the original and modified messages (i.e., $H(msg_0)$ and $H(msg_1)$) contain $\omega$ bytes with the same value at the same location (here $\omega$ is also called the number of \textit{hits}, which satisfies $0\leq \omega\leq g\equiv \lceil (n\times m)/8 \rceil$, and $W_N^\text{\,e}(0)+W_N^\text{\,e}(1)+\dots +W_N^\text{\,e}(g)=N$); and (2) the mean of the absolute difference per byte $\bar{d}_\text{byte}^\text{\,e}$ between $H(msg_0)$ and $H(msg_1)$ over $N$ draws. If the results of $W_N^\text{\,e}(\omega)$ and $\bar{d}_\text{byte}^\text{\,e}$ (the experimental values) are very close to their theoretical values, then the related hash function could be regarded as having a good property of collision resistance.

The number of hits on each draw can be obtained as follows: first, divide both $H(msg_0)$ and $H(msg_1)$ into $g$ bytes (if $n\times m$ is not divisible by 8, then add a prefix of $c=8-(n\times m)\bmod 8$ zeros to the hash values), so that the two hash values can be expressed as $h=e_1\lVert e_2\lVert e_3\lVert\dots\lVert e_g$ and $h'=e'_1\lVert e'_2\lVert e'_3\lVert\dots\lVert e'_g$, respectively ($e_j$ and $e'_j$ respectively represents the $j$th byte of $h$ and $h'$); second, compare $h$ and $h'$ byte by byte and compute $\omega$ according to
\begin{equation}\label{eq:9}
    \omega=\sum_{j=1}^g\delta\left[t(e_j),t(e'_j)\right],
\end{equation}
where $t(e_j)$ is the decimal value of $e_j$ and $\delta[\cdot]$ is the Kronecker delta function.

The theoretical value (denoted by $W_N^\text{\,t}(\omega)$) of $W_N^\text{\,e}(\omega)$ is given by the product of $N$ and the theoretical probability $P^\text{\,t}(\omega)$ that $\omega$ hits occur in $(h,h')$. Specifically, $P^\text{\,t}(\omega)$ is given by the binomial distribution formula
\begin{equation}\label{eq:10}
    P^\text{\,t}(\omega) =\frac{g!}{\omega !(g-\omega)!}\left(\frac{1}{2^8}\right)^{\omega} \left(1\!-\!\frac{1}{2^8}\right)^{g-\omega},
\end{equation}
and the theoretical number of draws with $\omega$ hits is obtained by
\begin{equation}\label{eq:11}
    W_N^\text{\,t}(\omega)=\text{int}\left[N\times P^\text{\,t}(\omega)\right],
\end{equation}
where $\text{int}[\cdot]$ denotes rounding a real number to its nearest integer.

Since $P^\text{\,t}\equiv \{P^\text{\,t}(\omega)\,|\,\omega =0,1,2,\dots g\}$ and $P^\text{\,e}\equiv\{W_N^\text{\,e}(\omega)/N\,|\,\omega =0,1,2,\dots g\}$ respectively describe the theoretical and experimental \textit{distributions of} $\omega$ (or simply \textit{hit distributions}), the similarity or difference between $\left\{W_N^\text{\,e}(\omega)\,|\,\omega =0,1,\dots g\right\}$ (hereafter, simply $\{W_N^\text{\,e}(\omega)\}$) and $\left\{W_N^\text{\,t}(\omega)\,|\,\omega =0,1,\dots g\right\}$ (hereafter, simply $\{W_N^\text{\,t}(\omega)\}$) could be measured by Kullback-Leibler divergence between $P^\text{\,t}$ and $P^\text{\,e}$, i.e.,
\begin{equation}
\begin{aligned}\label{eq:12}
    D_\text{KL}\left(P^\text{\,e}\lVert P^\text{\,t}\right) & =\sum_{\omega = 0}^g P^\text{\,e}(\omega)\text{log}_2\left(\frac{P^\text{\,e}(\omega)}{P^\text{\,t}(\omega)}\right) \\
    & =\sum_{\omega = 0}^g\frac{W_N^\text{\,e}(\omega)}{N}\text{log}_2\left(\frac{W_N^\text{\,e}(\omega)/N}{P^\text{\,t}(\omega)}\right);
\end{aligned}
\end{equation}
a smaller $D_\text{KL}\left(P^\text{\,e}\lVert P^\text{\,t}\right)$ indicates a closer similarity between $\{W_N^\text{\,e}(\omega)\}$ and $\{W_N^\text{\,t}(\omega)\}$.

The absolute difference per byte between $h$ and $h'$ is calculated by
\begin{equation}\label{eq:13}
    d_\text{byte}=\frac{1}{g}\sum_{j=1}^g\left|t(e_j)-t(e'_j)\right|,
\end{equation}
and the theoretical value (denoted by $\bar{d}_\text{byte}^\text{\,t}$ ) of the mean of $d_\text{byte}$ (denoted by $\bar{d}_\text{byte}^\text{\,e}$ ) over $N$ draws is $\bar{d}_\text{byte}^\text{\,t}=85.33$~\cite{Yang2021qwHash}.

The collision resistant test on QHFM-$L$ is performed with $N=10000$, and the simulation results are shown in Table~\ref{tab:table4}, where $W_N^\text{\,e}(4+)$ denotes the number of draws on which more than three hits occur in the hash values of the original and modified messages, that is, $W_N^\text{\,e}(4+)=N-[W_N^\text{\,e}(0)+W_N^\text{\,e}(1)+W_N^\text{\,e}(2)+W_N^\text{\,e}(3)]$. One may notice that the sums of $W_N^\text{\,t}(\omega)$ (over all $\omega$) for 264- and 128-bit hash schemes (or instances) are not equivalent to $N$, this is due to the rounding operations performed on $N\times P^\text{\,t}(\omega)$. The values of $D_\text{KL}\left(P^\text{\,e}\lVert P^\text{\,t}\right)$ and $\Delta\bar{d}_\text{byte}\equiv|\bar{d}_\text{byte}^\text{\,e}-\bar{d}_\text{byte}^\text{\,t}|$ for the existing schemes are deduced from the reported results of $\{W_N^\text{\,e}(\omega)\}$ and $\bar{d}_\text{byte}^\text{\,e}$ (or the mean of $d_\text{byte}\times g$), respectively.
\begin{table*}
\caption{Collision-resistance-test results for the proposed and existing QW-based hash functions}\label{tab:table4}
\begin{tabular}{lllcc}
\hline\noalign{\smallskip}
\textrm{Instances (schemes)} & $\qquad\{W_N^\text{\,e}(\omega)\}$ & $\qquad\{W_N^\text{\,t}(\omega)\}$ & $D_\text{KL}\left(P^\text{\,e}\lVert P^\text{\,t}\right)$ & $\Delta\bar{d}_\text{byte}$ \\
\hline\noalign{\smallskip}
    QHFM-296 & $8605,1312,81,2,0$ & $8652,1255,89,4,0$ & 0.000361 &  0.03 \\
    QHFM-264 & $8762,1159,74,5,0$ & $8788,1137,71,3,0$ & 0.000146 &  0.06 \\
    QHFM-221 & $8674,1230,93,3,0$ & $8962,984,52,2,0$ & 0.006711 &  2.48 \\
    QHFM-200 & $9071,895,34,0,0$ & $9068,889,42,1,0$ & 0.000302 &  0.03 \\
    QHFM-195 & $8066,1796,130,8,0$ & $9068,889,42,1,0$ & 0.069364 &  3.30 \\
    QHFM-136 & $9352,626,21,1,0$ & $9356,624,20,0,0$ & 0.000058 &  0.01 \\
    QHFM-120 & $9416,570,13,1,0$ & $9430,555,15,0,0$ & 0.000145 &  0.08 \\
    Yang21-296~\cite{Yang2021qwHash} & $8321,1547,110,22,0$ & $8652,1255,89,4,0$ & 0.008616  & 0.11 \\
    Yang19-264~\cite{Yang2019qwHash} & $9019,923,52,2,4$ & $8788,1137,71,3,0$ & 0.005647 &  4.43 \\
    Yang18-264~\cite{Yang2018qwHash264} & $8904,1026,68,2,0$ & $8788,1137,71,3,0$ & 0.000969 &  1.69 \\
    Yang18-221~\cite{Yang18qwHash221} & $9854,71,0,0,75$ & $8962,984,52,2,0$ & 0.188620 & N/A \\
    Li18-200~\cite{Li2018qwHash} & $8982,989,25,4,0$ & $9068,889,42,1,0$ & 0.001689 &  N/A \\
    Cao18-195~\cite{Cao2018qwHash} & $16063,314,6,0,0$ & $14856,1456,69,2,0$ & 0.066791 &  N/A \\
    Yang16-128~\cite{Yang2016qHash} & $9367,617,16,0,0$ & $9393,589,17,0,0$ & 0.000151 &  1.88 \\
\hline\noalign{\smallskip}
\end{tabular}
\end{table*}

The values of $D_\text{KL}\left(P^\text{\,e}\lVert P^\text{\,t}\right)$ indicate that the experimental result of hit distribution for QHFM-$L$ with $L\ge200$ has closer similarity to the theoretical distribution of $\omega$ than those for the existing ones with $L$-bit output length, and the Kullback-Leibler divergence between $P^\text{\,e}$ and $P^\text{\,t}$ for QHFM-$L$ with $L<200$ is on a par with that for its peer (Cao18-195 or Yang16-128). As for the average difference per byte in two hash values, the test results of $\bar{d}_\text{byte}$ for QHFM-296, QHFM-264, and QHFM-136 (together with QHFM-128) are closer to the theoretical value 85.33 than those for Yang21-296, Yang19-264, and Yang18-128, and the differences between $\bar{d}_\text{byte}^\text{\,e}$ and $\bar{d}_\text{byte}^\text{\,t}$ for the remaining instances of QHFM are very small. Therefore, the proposed hash scheme has a good capability of collision resistance.
\subsection{Resistance to Birthday Attacks}\label{sec:4.5}
Since the proposed hash function has variable digest length, one can easily obtain a QHFM instance that withstands birthday attacks by assigning appropriate values (large enough) to the parameters $m$ and $n$ according to the (cryptanalytic) hardware and software capabilities considered.

\section{Time and Space Complexity Analysis}\label{sec:5}
The hash value of an input message sent to a QW-based hash function can be calculated by cascading three stages: (1) initializing the state of the walker; (2) performing the underlying CAQW on a cycle according to the bit values of the message; and (3) calculating the hash value from the resulting probability distribution of the walker. The time and space complexity of the proposed scheme or an existing one can thus be obtained by analyzing the number of arithmetic operations taken by each stage of the related hashing process. Since the hash value is computed classically (quantum transforms are simulated by matrix multiplications), this section will concentrate on classical complexity.
\subsection{Time and space complexity of the proposed scheme}\label{sec:5.1}
The quantum state of the walker after $t$ steps of CQWM ($t\ge 0$) can be expressed as
\begin{equation}\label{eq:14}
    \Ket{\psi_t}=\sum_{x,j}A_t^{x,j}\Ket{x,j},
\end{equation}
where $A_t^{x,j} (x\in\mathbb{Z}_n,j\in\mathbb{Z}_8)$ is the amplitude of the 2-term basis state $\ket{x,j}$ at time $t$, and the correspondence between 2-term and 4-term basis states is described by equation group~(\ref{eq:5}). Before ($t=0$) and during ($t>0$) the walk, the state of the particle is identified with these $8n$ amplitudes.

When $t=0$, the particle is in the state $\Ket{\psi_0}=\text{cos}\alpha\Ket{0,2}+\text{sin}\alpha\Ket{0,3}$, which gives $A_0^{0,2}=\text{cos}\alpha$, $A_0^{0,3}=\text{sin}\alpha$, $A_0^{0, j}=0$ for $j\ne 2$ and $j\ne 3$, and $A_0^{x,j}=0$ for $x\ne 0$. Thus, the classical representation of the initial state $\Ket{\psi_0}$ can be specified using $8n$ assignments.

When $t>0$, if the $t$th message bit equals 0, the values of $\{A_t^{x,j}|x\in\mathbb{Z}_n,j\in\mathbb{Z}_8\}$ (hereafter, simply $\{A_t^{x,j}\}$) are determined by $U^{(0)}=S(I_n\otimes D^{(0)})(I_{4n}\otimes C^{(0)})$ and $\{A_{t-1}^{x,j}|x\in\mathbb{Z}_n,j\in\mathbb{Z}_8\}$ (hereafter, simply $\{A_{t-1}^{x,j}\}$). For the sake of simplicity of notation, we denote $C^{(0)}$ by $\bigl(\begin{smallmatrix} a_0 & b_0 \\ c_0 & d_0 \end{smallmatrix}\bigr)$, then the action of this coin operator on $\Ket{x,dr_2,dr_1,c}$ can be formulated as
\begin{equation}\label{eq:15}
\begin{aligned}
    C^{(0)}\!:\Ket{x,dr_2,dr_1,c}&\!\to\!\bar{c}(a_0\Ket{x,dr_2,dr_1,0}\!+\!c_0\Ket{x,dr_2,dr_1,1})+\\
                                   &c(b_0\Ket{x,dr_2,dr_1,0}\!+\!d_0\Ket{x,dr_2,dr_1,1}).
\end{aligned}
\end{equation}

Converting the 4-term states in expression~(\ref{eq:15}) into 2-term states gives
\begin{equation}\label{eq:16}
\begin{aligned}
    C^{(0)}:\Ket{x,j}\to & \left[(j \bmod 2)\oplus 1\right](a_0\Ket{x,j}+c_0\Ket{x,j+1})+ \\
                         & (j\bmod 2)(b_0\Ket{x,j-1}+d_0\Ket{x,j}),
\end{aligned}
\end{equation}
where $j+1$ and $j-1$ are both calculated using modular arithmetic under modulus 8. Combing expressions~(\ref{eq:16}), (\ref{eq:6}), and (\ref{eq:8}), one can obtain the action of $U^{(0)}$ on each 2-term basis state as well as on $\Ket{\psi_{t-1}}$ and then deduce the relation between $\{A_t^{x,j}\}$ and $\{A_{t-1}^{x,j}\}$. Specifically, the actions of $U^{(0)}$ on the components $A_{t-1}^{x,j}\Ket{x,j}$ of $\Ket{\psi_{t-1}}$ are
\begin{equation}\label{eq:17}
\begin{aligned}
    A_{t-1}^{x,0}\Ket{x,0} & \to a_0 A_{t-1}^{x,0}\Ket{x+1,4}+c_0 A_{t-1}^{x,0}\Ket{x-1,1},\\
    A_{t-1}^{x,1}\Ket{x,1} & \to b_0 A_{t-1}^{x,1}\Ket{x+1,4}+d_0 A_{t-1}^{x,1}\Ket{x-1,1},\\
    A_{t-1}^{x,2}\Ket{x,2} & \to a_0 A_{t-1}^{x,2}\Ket{x+1,6}+c_0 A_{t-1}^{x,2}\Ket{x-1,3},\\
    A_{t-1}^{x,3}\Ket{x,3} & \to b_0 A_{t-1}^{x,3}\Ket{x+1,6}+d_0 A_{t-1}^{x,3}\Ket{x-1,3},\\
    A_{t-1}^{x,4}\Ket{x,4} & \to a_0 A_{t-1}^{x,4}\Ket{x-1,0}+c_0 A_{t-1}^{x,4}\Ket{x+1,5},\\
    A_{t-1}^{x,5}\Ket{x,5} & \to b_0 A_{t-1}^{x,5}\Ket{x-1,0}+d_0 A_{t-1}^{x,5}\Ket{x+1,5},\\
    A_{t-1}^{x,6}\Ket{x,6} & \to a_0 A_{t-1}^{x,6}\Ket{x-1,2}+c_0 A_{t-1}^{x,6}\Ket{x+1,7},\\
    A_{t-1}^{x,7}\Ket{x,7} & \to b_0 A_{t-1}^{x,7}\Ket{x-1,2}+d_0 A_{t-1}^{x,7}\Ket{x+1,7},
\end{aligned}
\end{equation}
where $x\pm 1$ are calculated using modular arithmetic under modulus $n$. Summing up the transformed results on the right side, one can observe that the amplitudes of the walker being at position $x$ at time $t-1$ contribute $a_0A_{t-1}^{x,0}+b_0A_{t-1}^{x,1}$ to $A_t^{x+1,4}$, $c_0A_{t-1}^{x,0}+d_0A_{t-1}^{x,1}$ to $A_t^{x-1,1}$, and $a_0A_{t-1}^{x,2}+b_0A_{t-1}^{x,3}$ to $A_t^{x+1,6}$, etc.; here $A_t^{x\pm 1,j}$ is the amplitude of $\Ket{x\pm 1\pmod{n},j}$ at time $t$. Moreover, the amplitudes of being at position $x$ at time $t-1$ only contribute to $A_t^{x-1,j}$ with $j\leq 3$ and to $A_t^{x+1,j}$ with $j\ge 4$; conversely, the former 4 amplitudes (with $j\leq 3$) at an arbitrary position $x$ are contributed by the amplitudes at position $x+1\pmod{n}$, while the latter 4 amplitudes (with $j\ge 4$) at position $x$ are contributed by those at position $x-1\pmod{n}$. As a result, each amplitude of being at position $x$ at time $t$, denoted by $A_t^{x,j}$, is only contributed by the amplitudes of being at a single position ($x+1\pmod{n}$ or $x-1\pmod{n}$) at time $t-1$. Thus, the relation between $\{A_t^{x,j}\}$ and $\{A_{t-1}^{x,j}\}$ after a step of QW1M can be expressed as follows:
\begin{equation}\label{eq:18}
\begin{aligned}
    A_t^{x,0}=a_0 A_{t-1}^{x+1,4}+b_0 A_{t-1}^{x+1,5}, A_t^{x,1}=c_0 A_{t-1}^{x+1,0}+d_0 A_{t-1}^{x+1,1},\\
    A_t^{x,2}=a_0 A_{t-1}^{x+1,6}+b_0 A_{t-1}^{x+1,7}, A_t^{x,3}=c_0 A_{t-1}^{x+1,2}+d_0 A_{t-1}^{x+1,3},\\
    A_t^{x,4}=a_0 A_{t-1}^{x-1,0}+b_0 A_{t-1}^{x-1,1}, A_t^{x,5}=c_0 A_{t-1}^{x-1,4}+d_0 A_{t-1}^{x-1,5},\\
    A_t^{x,6}=a_0 A_{t-1}^{x-1,2}+b_0 A_{t-1}^{x-1,3}, A_t^{x,7}=c_0 A_{t-1}^{x-1,6}+d_0 A_{t-1}^{x-1,7}.
\end{aligned}
\end{equation}

Analogously, if the $t$th message bit equals 1, the values of $\{A_t^{x,j}\}$ are determined by $U^{(1)}=S(I_n\otimes D^{(1)})(I_{4n}\otimes C^{(1)})$ and $\{A_{t-1}^{x,j}\}$. We denote $C^{(1)}$ by $\bigl(\begin{smallmatrix} a_1 & b_1 \\ c_1 & d_1 \end{smallmatrix}\bigr)$, then the action of $C^{(1)}$ on $\Ket{x,j}$ is
\begin{equation}\label{eq:19}
\begin{aligned}
    C^{(1)}:\Ket{x,j}\to & \left[(j \bmod 2)\oplus 1\right](a_1\Ket{x,j}+c_1\Ket{x,j+1})+ \\
                         & (j\bmod 2)(b_1\Ket{x,j-1}+d_1\Ket{x,j}).
\end{aligned}
\end{equation}

A Combination of expressions~(\ref{eq:19}), (\ref{eq:7}), and (\ref{eq:8}) gives the actions of $U^{(1)}$ on the components of $\Ket{\psi_{t-1}}$:
\begin{equation}\label{eq:20}
\begin{aligned}
    A_{t-1}^{x,0}\Ket{x,0} & \to a_1 A_{t-1}^{x,0}\Ket{x+1,4}+c_1 A_{t-1}^{x,0}\Ket{x-1,1},\\
    A_{t-1}^{x,1}\Ket{x,1} & \to b_1 A_{t-1}^{x,1}\Ket{x+1,4}+d_1 A_{t-1}^{x,1}\Ket{x-1,1},\\
    A_{t-1}^{x,2}\Ket{x,2} & \to a_1 A_{t-1}^{x,2}\Ket{x-1,0}+c_1 A_{t-1}^{x,2}\Ket{x+1,5},\\
    A_{t-1}^{x,3}\Ket{x,3} & \to b_1 A_{t-1}^{x,3}\Ket{x-1,0}+d_1 A_{t-1}^{x,3}\Ket{x+1,5},\\
    A_{t-1}^{x,4}\Ket{x,4} & \to a_1 A_{t-1}^{x,4}\Ket{x+1,6}+c_1 A_{t-1}^{x,4}\Ket{x-1,3},\\
    A_{t-1}^{x,5}\Ket{x,5} & \to b_1 A_{t-1}^{x,5}\Ket{x+1,6}+d_1 A_{t-1}^{x,5}\Ket{x-1,3},\\
    A_{t-1}^{x,6}\Ket{x,6} & \to a_1 A_{t-1}^{x,6}\Ket{x-1,2}+c_1 A_{t-1}^{x,6}\Ket{x+1,7},\\
    A_{t-1}^{x,7}\Ket{x,7} & \to b_1 A_{t-1}^{x,7}\Ket{x-1,2}+d_1 A_{t-1}^{x,7}\Ket{x+1,7}.
\end{aligned}
\end{equation}
Thus, the relation between $\{A_t^{x,j}\}$ and $\{A_{t-1}^{x,j}\}$ after a step of QW2M can be expressed as
\begin{equation}\label{eq:21}
\begin{aligned}
    A_t^{x,0}=a_1 A_{t-1}^{x+1,2}+b_1 A_{t-1}^{x+1,3}, A_t^{x,1}=c_1 A_{t-1}^{x+1,0}+d_1 A_{t-1}^{x+1,1},\\
    A_t^{x,2}=a_1 A_{t-1}^{x+1,6}+b_1 A_{t-1}^{x+1,7}, A_t^{x,3}=c_1 A_{t-1}^{x+1,4}+d_1 A_{t-1}^{x+1,5},\\
    A_t^{x,4}=a_1 A_{t-1}^{x-1,0}+b_1 A_{t-1}^{x-1,1}, A_t^{x,5}=c_1 A_{t-1}^{x-1,2}+d_1 A_{t-1}^{x-1,3},\\
    A_t^{x,6}=a_1 A_{t-1}^{x-1,4}+b_1 A_{t-1}^{x-1,5}, A_t^{x,7}=c_1 A_{t-1}^{x-1,6}+d_1 A_{t-1}^{x-1,7}.
\end{aligned}
\end{equation}

Relations~(\ref{eq:18}) and (\ref{eq:21}) show that, given the amplitudes $\{A_{t-1}^{x_0,j}|j\in\mathbb{Z}_8\}$ of being at a fixed position $x_0$ at time $t-1$, the values of the amplitudes $\{A_t^{x_0,j}|j\in\mathbb{Z}_8\}$ of being at $x_0$ at time $t$ can be calculated using 16 multiplications and 8 additions, which means all amplitudes at each time step of CQWM on a cycle with $n$ nodes can be obtained using $O(n)$ basic arithmetic operations. To perform these operations, one needs to store the old (or the initial) $8n$ amplitudes and their 8 possible coefficients ($a_1, a_2,\dots, d_1, d_2$) to calculate the new $8n$ amplitudes, and the values of both old and new amplitudes are refreshed at each time step. If the input message $msg$ is a binary string of $M$ bits, then the values of $\{A_M^{x,j}\}$ can be obtained using $O(Mn)$ basic operations with $O(n)$ memory space. Finally, the hash value is computed from $\{A_M^{x,j}\}$ using $O(n)$ multiplications and $O(n)$ modulo operations with $O(n)$ space. Thus, the time and space complexity of QHFM with input length $M$ are $O(Mn)$ and $O(n)$, respectively.

In particular, if one wants to obtain an $L$-bit hash value ($L$ is a multiple of $m$) of $msg$ using QHFM-$L$, then the cycle utilized by QHFM-$L$ has $L/m=O(L)$ nodes (here $m$ is constant with respect to the input length $M$); in this case, the hash value is produced with $O(ML)$ time and $O(L)$ space.
\subsection{Time and space complexity comparison of QW-based hash schemes}\label{sec:5.2}
In a similar way, one can deduce the time and space complexity of the existing QW-based hash functions~\cite{Yang2021qwHash,Yang2019qwHash,Yang2018qwHash264,Yang18qwHash221,Li2018qwHash,Cao2018qwHash,Yang2016qHash,Li2013qwHash} with respect to the input and output length. To facilitate discussion, we divide the existing schemes into four groups: (1) the hash functions based on one-dimensional one-particle quantum walks ~\cite{Yang2021qwHash,Yang2019qwHash,Yang2018qwHash264,Yang18qwHash221}, (2) the hash function based on two-dimensional one-particle quantum walks~\cite{Li2018qwHash}, (3) the hash function based on quantum walks on Johnson graphs~\cite{Cao2018qwHash}, and (4) the hash functions based on one-dimensional two-particle quantum walks ~\cite{Yang2016qHash,Li2013qwHash}.

Again, suppose all schemes produce hash values of length $L$. In this case, the schemes in group (1) utilize a cycle with $O(L)$ nodes, and the amplitudes of the particle being at each node at time $t$ can be calculated from the amplitudes of being at the two neighbors of this node at time $t-1$ (possibly calculated from the amplitudes of being at a single neighbor or remain unchanged during broken-line quantum walks~\cite{Yang2021qwHash}) using constant number of (basic arithmetic) operations. In group (2), Li18-200 utilizes cycles of length $O(\sqrt{L})$ in two-dimensional space, which lead to $O(L)$ positions for the walker, and the amplitudes of being at position $(x,y)$ at time $t$ can be calculated from the amplitudes of being at $(x\pm 1,y\pm 1)$ at time $t-1$ using constant number of operations. In group (3), Cao18-195 utilizes a Johnson graph $J(n,1)$ with $n=O(L)$ nodes, and the amplitudes of the particle being at each node at time $t$ can be calculated from the amplitudes of being at the remaining $n-1$ nodes at time $t-1$ using $O(L)$ operations. Similar to group (2), schemes in group (4) also utilize a cycle with $O(\sqrt{N})$ nodes, which leads to $O(L)$ position pairs for the two particles, and the amplitudes of the first and second particles being respectively at nodes $x$ and $y$ at time $t$ can be calculated from the amplitudes of the two particles being respectively at $x\pm 1$ and $y\pm 1$ at time $t-1$ using constant number of operations.

Thus, except for Cao18-195~\cite{Cao2018qwHash}, which performs $O(L^2)$ operations at each time step, the existing QW-based schemes ~\cite{Yang2021qwHash,Yang2019qwHash,Yang2018qwHash264,Yang18qwHash221,Li2018qwHash,Yang2016qHash,Li2013qwHash} take $O(L)$ time to calculate the amplitudes at time $t$ from the amplitudes at time $t-1$. If the input message is of bit-length $M$, then the resulting amplitudes at time $M$ can be obtained with $O(ML^2)$ and $O(ML)$ operations in Cao18-195 and the remaining schemes, respectively. After that, for all these schemes ~\cite{Yang2021qwHash,Yang2019qwHash,Yang2018qwHash264,Yang18qwHash221,Li2018qwHash,Cao2018qwHash,Yang2016qHash,Li2013qwHash}, the hash value is computed from the resulting $O(L)$ amplitudes with $O(L)$ operations. Therefore, the time complexities of Cao18-195 and the other QW-based hash schemes are $O(ML^2)$ and $O(ML)$, respectively. Since each amplitude at time $t$ is a linear combination of the amplitudes of $O(1)$ or $O(L)$ positions at time $t-1$ in all schemes, and the resulting probability distribution takes $O(L)$ space as well, the space complexities of the existing schemes all equals $O(L)$.

As a result, the proposed scheme has the same time and space complexity as the existing QW-based hash schemes except Cao18-195, whose time complexity is slightly greater than that of the other schemes, including the proposed one.

\section{Conclusion}\label{sec:6}
In this paper, a new hash function QHFM based on quantum walks with one- and two-step memory on circles is constructed, whose statistical properties as well as time and space complexity are evaluated and compared with the existing QW-based hash functions.

Unlike the existing analyses of hash schemes based on quantum walks without memory, where a single indicator $\overline{T}$ is used to evaluate the uniform distribution property, we adopted an additional indicator $\Delta Q$ to assess this property, since $\overline{T}$ alone is closely related to $P$, implying that it also suggests the diffusion and confusion properties. In the collision resistance analysis, we use Kullback-Leibler divergence to evaluate the similarity between the experimental and theoretical distributions of $\omega$, so that the difference between $\{W_N^\text{\,e}(\omega)\}$ and $\{W_N^\text{\,t}(\omega)\}$ can be indicated by a single number.

The analysis results show that QHFM has near-ideal statistical performance and takes no more time and space than its peers, and they also suggest that alternately running two quantum walks differing in more than one respects, including coin operator and memory length, can also yield good hash functions. Thus, it is unnecessary to restrict the component parts of a controlled alternate quantum walk to a single kind of walk (equipped with controlled coins). In future work, we will explore the possibility of combining two quantum walks with more differences and investigate the effect of those differences on the performance of the resulting hash function.
\begin{acknowledgements}
The authors gratefully acknowledge the financial support from the China Postdoctoral Science Foundation under Grant No. 2021M691148, the Hubei Provincial Science and Technology Major Project of China under Grant No. 2020AEA011, and the Key Research $\&$ Development Plan of Hubei Province of China under Grant No. 2020BAB100.
\end{acknowledgements}


%
\end{document}